# Hyperbolic plasmons on uniaxial metamaterials

Osamu Takayama,[a)] and Andrei V. Lavrinenko

*DTU Fotonik—Department of Photonics Engineering, Technical University of Denmark, DK–2800 Kongens Lyngby, Denmark*

We analyze surface electromagnetic waves with hyperbolic dispersion supported at the interface between a semi–infinite isotropic medium and an effective uniaxial material. Apart from known types plasmons with hyperbolic dispersion curve, sometimes referred to as Dyakonov plasmons [Z. Jacob and E. E. Narimanov, APL **93**, 221109 (2008)], we classify two new types of surface waves with hyperbolic dispersion. One type of such waves, in contrary to Dyakonov plasmons, does not require hyperbolic metamaterials to be involved. These hybrid–polarized plasmon modes with both TE and TM electromagnetic components are directional. Their propagation direction can be controlled by changing material parameters.

**Main Text**

A surface electromagnetic wave or surface wave (SW) is supported at an interface between two dissimilar media and propagates along the interface. Its field amplitude exponentially decays away from the boundary.[1] The research on surface waves has been flourishing in the last decade thanks to their unique properties of surface sensitivity and field localization, which results in applications in sensing, light–trapping, or imaging based on the near–field techniques, thus contributing to the nanophotonics. The most known example of optical surface waves is a surface plasmon polariton, a TM–polarized surface wave formed on the interfaces between metals and dielectrics.[2-11] Another example of a SW is a Dyakonov surface wave, a highly directional hybrid SW supported at the interface between two transparent dielectrics, where at least one of them is a birefringent medium. Dyakonov SWs have recently been emerging as a lossless complement to plasmons,[12-16] as well as subdiffraction confinement of light by transparent anisotropic media.[17,18]

Apart from plasmons and Dyakonov SWs, a new kind of SWs, sometimes referred to as Dyakonov plasmons[19] on highly anisotropic metamaterials, named hyperbolic metamaterials[20] with at least one of its axial permittivity negative, gaining attention due to its unique hyperbolic angular dispersion.[21-26] The effective refractive index, $N$, of the Dyakonov plasmons increases exponentially to infinity with the angle, $\theta$, measured from the optical axis (OA) of the birefringent media in the plane of the interface (Fig.1). Extremely large values of the effective refractive index mean high localization of SWs. So, the hyperbolic angular dispersion allows us to control the degree of field localization by changing the propagation angle of the plasmons in the plane of the interface. Moreover, directional SWs enable us to perform steering operations of light on the surface,[11,16,24] and obtain directional emission from a light source such as a quantum dot.[19,23] In the same line, the possibility

[a)]Electronic mail: otak@fotonik.dtu.dk

of engineering the effective permittivities and permeabilities in photonic metamaterials opens new perspectives in SWs.[15,27-29] In particular, understanding the physics of surface waves in terms of permittivities of materials and their influence on the dispersion is of the topical importance.

In this Letter, we systematically investigate *hyperbolic plasmons*, *e.g.* plasmons with hyperbolic angular dispersion existing on the interface between isotropic and uniaxial anisotropic materials by taking all possible combinations of their permittivities into account. Our classification scheme helps to identify two new types of hyperbolic plasmons with unusual dispersion properties, which require some certain values of material parameters. We also clarify the existence conditions for each hyperbolic plasmon and show that they do not necessarily require metamaterials that possess hyperbolic dispersion.

We consider a planar interface between two non–magnetic semi–infinite media, an isotropic and uniaxial one. The isotropic medium is characterized by permittivity $\varepsilon_c$, and the uniaxial material is characterized by the effective ordinary and extraordinary permittivities, $\varepsilon_o$ and $\varepsilon_e$, respectively. The OA of the uniaxial material lies in the *y*, *z*–interface plane, forming angle $\theta$ with the propagation direction *z* as illustrated in Fig. 1. The *x*–axis is perpendicular to the interface. Assuming harmonic surface electromagnetic plane waves with the pure imaginary wavevector component in the *x* direction, the eigenvalue equation of surface waves at the isotropic–uniaxial boundary can be expressed by,[28]

$$\varepsilon_o A_e B_o \sin^2\theta + \gamma_o^2 A_o B_e \cos^2\theta = 0 \tag{1}$$

where

$$A_{o,e} = \gamma_{o,e} + \gamma_c, B_o = \varepsilon_c \gamma_o + \varepsilon_o \gamma_c,$$

$$B_e = \varepsilon_c \gamma_o + \varepsilon_o \gamma_c \gamma_e / \gamma_o, \gamma_c = (N^2 - \varepsilon_c)^{1/2},$$

$$\gamma_o = (N^2 - \varepsilon_o)^{1/2}, \gamma_e(\theta) = \varepsilon_e / \varepsilon_e(\theta)(N^2 - \varepsilon_e(\theta))^{1/2}, \tag{2}$$

and

$$\varepsilon_e(\theta) = \varepsilon_o \varepsilon_e / (\varepsilon_o \sin^2\theta + \varepsilon_e \cos^2\theta) \quad (0° \leq \theta \leq 90°). \tag{3}$$

Eq. (1) is a general expression for lossless surface wave solutions supported at the isotropic–uniaxial interface between two media independently of the sign of the permittivities. Since effective refractive index $N$ for hyperbolic plasmons increases to infinity for certain cutoff angle, $\theta_{min,max}$, we have from Eq. (1) for $N \to \infty$,

$$\varepsilon_c = -\varepsilon_o \sqrt{\frac{\varepsilon_e}{\varepsilon_e(\theta)}} \qquad (4)$$

For Eq. (4) to hold, $\varepsilon_e/\varepsilon_e(\theta) > 0$ and the sign of $\varepsilon_o$ and $\varepsilon_c$ must be opposite. Thus, we can deduce that for hyperbolic plasmons to exist, it is necessary that signs of the materials permittivities form one of the following combinations:

$$(\varepsilon_o, \varepsilon_e \mid \varepsilon_c) = (+,+\mid -), (-,+\mid +), (+,-\mid -), and\ (-,-\mid +). \qquad (5)$$

We will refer to these groups as dielectric metrics. Therefore, among eight possible metrics for materials in contact (TABLE I), we identify four cases belonging to hyperbolic plasmons [cases (i) to (iv)], as well as Dyakonov SWs that occur for $(\varepsilon_o, \varepsilon_e \mid \varepsilon_c) = (+, + \mid +)$.[12] In order to classify the types of surface waves and their existence conditions, TABLE I summarizes the requirements for each type of SWs in terms of permittivities. Case (i) contains both three types of hyperbolic plasmons and three types of plasmons with elliptic dispersions,[30] which we refer to as *elliptic plasmons*. Case (ii) has two types of hyperbolic plasmons that propagates either from $\theta = \theta_{min}$ to $\theta_{max} = 90°$, or $\theta = \theta_{min}$ to $\theta = \theta_{max}$, ($\neq 90°$), where $\theta_{min}$ and $\theta_{max}$ ($\theta_{min} < \theta_{max}$) are the lower and upper cutoff angles, respectively. Case (ii) has been investigated previously.[19,21,22] Cases (iii) and (iv) have one and two forms of hyperbolic plasmons, respectively, as well as elliptic plasmons. In this paper, we restrict ourselves to the unexplored hyperbolic plasmons of cases (iii) and (iv).

For case (iii) with metrics $(\varepsilon_o, \varepsilon_e \mid \varepsilon_c) = (+, - \mid -)$, from Eq. (4) $\varepsilon_e/\varepsilon_e(\theta) \geq 1$, so $(\varepsilon_e/\varepsilon_e(\theta))^{1/2} \geq 1$. Hence, the condition for the hyperbolic plasmon to exist is

$$|\varepsilon_c| \leq \varepsilon_o. \qquad (6)$$

Note that in the case where $|\varepsilon_c| > \varepsilon_o$, the plasmon is no longer hyperbolic, it has elliptic dispersion. The lower cutoff angle $\theta_{min}$ is where $\varepsilon_e(\theta) = 0$, so from Eq. (3) we have

$$\theta_{min} = \sin^{-1} \sqrt{\frac{\varepsilon_e}{\varepsilon_e - \varepsilon_o}}. \qquad (7)$$

The upper cutoff angle, $\theta_{max}$, at which $N \to \infty$, occurs, as it follows from Eq. (4), at

$$\theta_{max} = \sin^{-1}\sqrt{\frac{\varepsilon_c^2 - \varepsilon_o\varepsilon_e}{\varepsilon_o(\varepsilon_o - \varepsilon_e)}}. \tag{8}$$

We see from Eq. (8) that in the particular case of $|\varepsilon_c| = \varepsilon_o$, $\theta_{max} = 90°$.

We now present numerical results illustrating the dispersion feature of case (iii), as well as their polarization characteristics. Fig. 2(a) shows the dispersion in the Fourier space. Without loss of generality, we specified parameters for calculation of each case; $|\varepsilon_c| < \varepsilon_o$: $(\varepsilon_o, \varepsilon_e | \varepsilon_c) = (2, -1 | -1)$, $|\varepsilon_c| = \varepsilon_o$ : $(2, -1 | -2)$, and $|\varepsilon_c| > \varepsilon_o$: $(2, -1 | -10)$. Note that the lower cutoff angle $\theta_{min}$ from Eq. (7) applies for all cases. These hyperbolic plasmons show unconventional hyperbolic dispersion with narrower angular cone than that of case (i, ii, iv) and significant increase in effective refractive index $N$ [see $|\varepsilon_c| = \varepsilon_o$ case in Fig.2 (a)]. When $|\varepsilon_c| > \varepsilon_o$, the plasmon shows elliptic dispersion [see Fig. 2(a)]. Fig. 2(b) shows the ratio of electric field components along the $y$ and $x$–axis at the interface ($x = 0$), representing the measure of polarization hybridity at angle $\theta$. It is obvious that the hyperbolic plasmons and elliptic plasmons are TM–dominant waves. At the upper cutoff angle, $\theta_{max}$, the polarization becomes the same as for the TM–polarized wave.

Case (iv) with metric $(\varepsilon_o, \varepsilon_e | \varepsilon_c) = (-, - | +)$ embraces two forms of hyperbolic plasmons with permittivities: $|\varepsilon_o| > |\varepsilon_e|$ and $|\varepsilon_o| < |\varepsilon_e|$. In the former case with $|\varepsilon_o| > |\varepsilon_e|$, from Eq. (4)

$$(\varepsilon_o\varepsilon_e)^{1/2} \leq \varepsilon_c \leq |\varepsilon_o| \tag{9}$$

gives the condition for hyperbolic dispersion. In the case of $\varepsilon_c < (\varepsilon_o\varepsilon_e)^{1/2}$, the plasmon exhibits elliptic dispersion, and $\varepsilon_c > |\varepsilon_o|$ case attributes no surface wave. The numerical results are presented in Fig. 3(a) with parameters; $(\varepsilon_o\varepsilon_e)^{1/2} < \varepsilon_c < |\varepsilon_o|$: $(\varepsilon_o, \varepsilon_e | \varepsilon_c) = (-10, -1 | 5)$, $\varepsilon_c = (\varepsilon_o\varepsilon_e)^{1/2}$: $(-10, -0.1 | 1)$, and $\varepsilon_c < (\varepsilon_o\varepsilon_e)^{1/2}$: $(-10, -1 | 1)$. It is important to emphasize that the dispersion law of plasmons in Eq. (9) case is similar to the dispersion law in case (ii), Dyakonov plasmons with hyperbolic metamaterials,[19] although the present material is not a hyperbolic metamaterial. Eq. (8) gives the lower cutoff angle $\theta_{min}$, for hyperbolic plasmons. In the limited case of $\varepsilon_c = (\varepsilon_o\varepsilon_e)^{1/2}$, $N \to \infty$ at $\theta_{min} = 0°$ as it can be seen from the Fig. 3 (a). Once $\varepsilon_c < (\varepsilon_o\varepsilon_e)^{1/2}$ is fulfilled, the dispersion of plasmons drastically changes from hyperbolic to elliptic. These hyperbolic plasmons are TE–dominant as shown in Fig. 3(c) in contrast to conventional plasmons that are the TM waves. Such hybrid polarization of SWs enables us to couple to SWs irrespectively of the input polarization.[14,16]

Similarly, for the case of $|\varepsilon_o|<|\varepsilon_e|$ the condition for a hyperbolic plasmon is

$$|\varepsilon_o| \le \varepsilon_c \le (\varepsilon_o \varepsilon_e)^{1/2}. \qquad (10)$$

When $\varepsilon_c < |\varepsilon_o|$, the plasmons have elliptic dispersion. $\varepsilon_c > (\varepsilon_o\varepsilon_e)^{1/2}$ gives no solution. The parameters used for the calculations in Fig. 3 (b) are: $|\varepsilon_o|< \varepsilon_c < (\varepsilon_o\varepsilon_e)^{1/2}$: $(\varepsilon_o, \varepsilon_e | \varepsilon_c) = (-1, -10 | 2)$, $\varepsilon_c = |\varepsilon_o|$: $(-1, -10 | 1)$, and $\varepsilon_c < |\varepsilon_o|$: $(-1.5, -10 | 1)$. Their dispersions are similar to those of the $|\varepsilon_o|>|\varepsilon_e|$ case but rotated by 90° from the optical axis. The upper cutoff angle $\theta_{max}$ for hyperbolic plasmons is given by Eq. (8). In the limited case of $\varepsilon_c = |\varepsilon_o|$, $N \rightarrow \infty$ at $\theta_{max} = 90°$, see Fig. 3 (b). The polarization of these plasmons is also hybrid [see Fig. 3 (d)], but they are more TM–dominant than the case of $|\varepsilon_o|>|\varepsilon_e|$ setting. From Fig. 3 (c) and (d) we can see that when a hyperbolic plasmon propagates at lower and upper cutoff angle $\theta = \theta_{min,max}$, the plasmon becomes a TM–polarized wave. Note that for $|\varepsilon_o|=|\varepsilon_e|$, the uniaxial material convert to isotropic one and supports plasmons with the circular dispersion.

To summarize, we have systematically studied the existence of surface waves at the interface between isotropic and uniaxial media based on the signs of their permittivities. We discover that two new types of hyperbolic plasmons with unique dispersions exist even though the uniaxial anisotropic material itself does not possess the hyperbolic dispersion. Such new surface wave solutions originate from the anisotropic permittivities of the uniaxial media, resulting in unique hyperbolic–like wavevector dependencies. We also found that plasmons with elliptic dispersion can exist under certain conditions. Importantly, we have shown that the dispersion of these surface waves can be switched (from hyperbolic to elliptic and back) and also their directionality can be varied by changing material parameters, leading to switching, routing, and directional emission of light at nanoscale.

This work was supported by Villum Fonden. Authors thank Sergei Zhukovsky and Radu Malureanu for fruitful discussions.

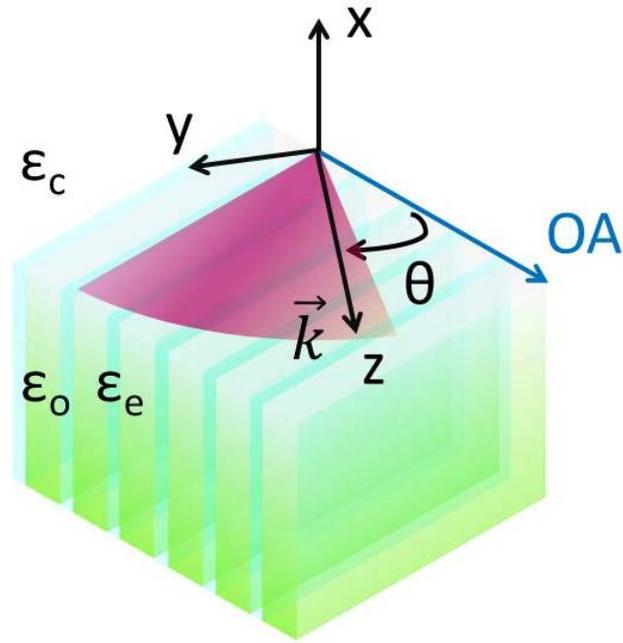

FIG. 1. (Color online) Geometry under consideration. The interface between the isotropic layer with permittivity $\varepsilon_c$ and uniaxial birefringent metamaterial characterized by ordinary and extraordinary permittivities, $\varepsilon_o$ and $\varepsilon_e$, respectively. The optical axis (OA) is in the plane of the interface. A surface wave propagates along the $z$ direction which forms angle $\theta$ with the OA.

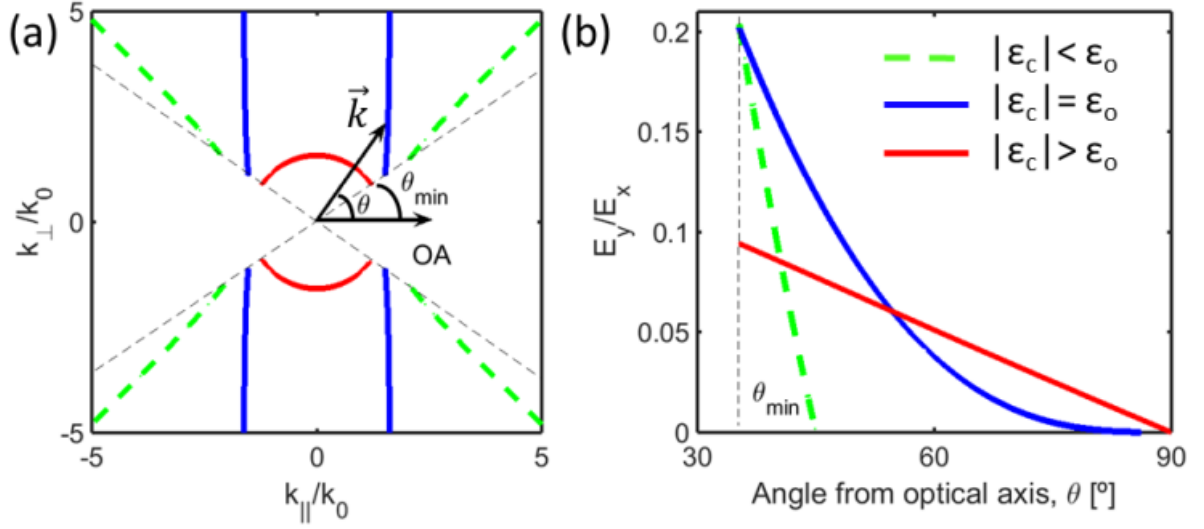

FIG. 2. (Color online) (a) Normalized wavevector of the plasmon modes for case (iv) with metric $(\varepsilon_o, \varepsilon_e | \varepsilon_c) = (+, - | -)$. (b) Fields ratio $E_y/E_x$ at the interface ($x = 0$). The black dotted line shows the lower cutoff angle, $\theta_{min}$, of the plasmons. The parameters used for calculation are $|\varepsilon_c| < \varepsilon_o$: $(\varepsilon_o, \varepsilon_e | \varepsilon_c) = (2, -1 | -1)$, $|\varepsilon_c| = \varepsilon_o$: $(2, -1 | -2)$, and $|\varepsilon_c| > \varepsilon_o$: $(2, -1 | -10)$. Lower cutoff angle $\theta_{min}$ applies for all cases. The color scheme is the same in both (a) and (b) figures.

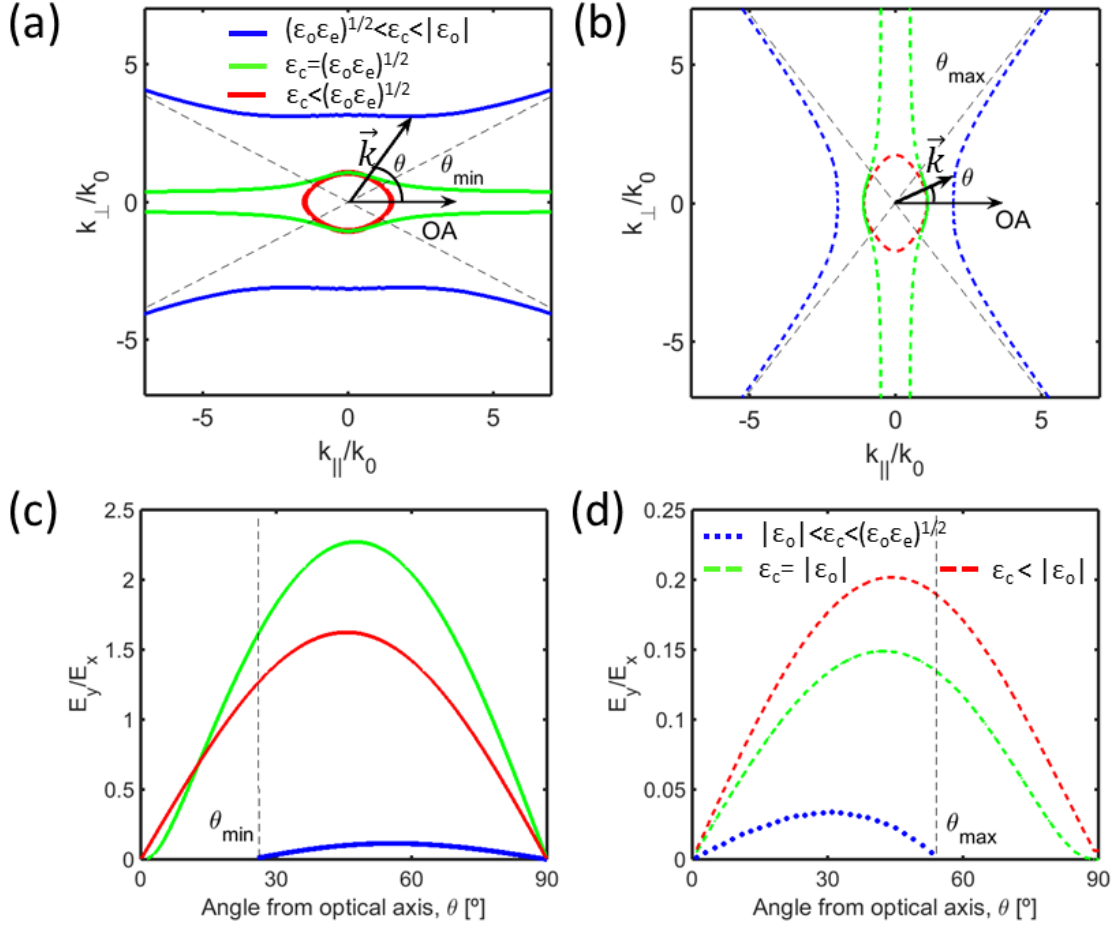

FIG. 3. (Color online) Normalized wavevector of the plasmon modes for case (iv) with metric $(\varepsilon_o, \varepsilon_e|\varepsilon_c) = (-, -\,|\,+)$ for (a) $|\varepsilon_o|>|\varepsilon_e|$ and (b) $|\varepsilon_o|<|\varepsilon_e|$. Fields ratio $E_y/E_x$ at $x=0$ for (c) $|\varepsilon_o|>|\varepsilon_e|$ and (d) $|\varepsilon_o|<|\varepsilon_e|$. The parameters used for calculation in (a) and (c) are $(\varepsilon_o\varepsilon_e)^{1/2} < \varepsilon_c <|\varepsilon_o|$: $(\varepsilon_o, \varepsilon_e\,|\,\varepsilon_c) = (-10, -1\,|\,5)$, $\varepsilon_c = (\varepsilon_o\varepsilon_e)^{1/2}$: $(-10, -0.1\,|\,1)$, and $\varepsilon_c < (\varepsilon_o\varepsilon_e)^{1/2}$: $(-10, -1\,|\,1)$. For (b) and (d), $|\varepsilon_o|< \varepsilon_c < (\varepsilon_o\varepsilon_e)^{1/2}$: $(\varepsilon_o, \varepsilon_e\,|\,\varepsilon_c) = (-1, -10\,|\,2)$, $\varepsilon_c = |\varepsilon_o|$: $(-1, -10\,|\,1)$, and $\varepsilon_c < |\varepsilon_o|$: $(-1.5, -10\,|\,1)$. Note that lower and cutoff angle $\theta_{min,max}$ only apply for $(\varepsilon_o\varepsilon_e)^{1/2} < \varepsilon_c <|\varepsilon_o|$ and $|\varepsilon_o|< \varepsilon_c < (\varepsilon_o\varepsilon_e)^{1/2}$ cases, respectively.

TABLE I. Classification and conditions of surface waves on isotropic–uniaxial interface.

| Uniaxial medium | | Isotropic medium | Surface waves and their existence conditions |
|---|---|---|---|
| $\varepsilon_o$ | $\varepsilon_e$ | $\varepsilon_c$ | |
| + | + | + | Dyakonov surface waves, $\varepsilon_o < \varepsilon_c < \varepsilon_e$ [12] |
| | | — | case (i) 3 Hyperbolic plasmons. For the conditions, see Ref.30[a] |
| — | + | + | case (ii) 2 Hyperbolic plasmons (Dyakonov plasmons), $0 < \varepsilon_c \leq |\varepsilon_o|\varepsilon_e/|\varepsilon_e - \varepsilon_o|$, and $|\varepsilon_o|\varepsilon_e/|\varepsilon_e - \varepsilon_o| < \varepsilon_c < |\varepsilon_o|$ [19,21] |
| | | — | No surface wave |
| + | — | + | No surface wave |
| | | — | case (iii) 1 Hyperbolic plasmons, $|\varepsilon_c| \leq \varepsilon_o$, and 1 elliptic plasmon, $|\varepsilon_c| > \varepsilon_o$ |
| — | — | + | case (iv) 2 Hyperbolic plasmons, $(\varepsilon_o\varepsilon_e)^{1/2} \leq \varepsilon_c < |\varepsilon_o|$ for $|\varepsilon_o| > |\varepsilon_e|$, $|\varepsilon_o| \leq \varepsilon_c < (\varepsilon_o\varepsilon_e)^{1/2}$ for $|\varepsilon_o| < |\varepsilon_e|$. 2 elliptic plasmon, $\varepsilon_c < (\varepsilon_o\varepsilon_e)^{1/2}$ and $\varepsilon_c < |\varepsilon_o|$ |
| | | — | No surface wave |

[a]case (i) also embraces 3 elliptic plasmons.